\documentclass{www2007-submission}
\usepackage{subfigure}
\usepackage{graphicx}

\newtheorem{example}{Example}
\usepackage{url} 
\usepackage{listings}
\usepackage{color}  
\usepackage{ifpdf}

 \graphicspath{{../examples/}{icfsh10-200/}}

\begin{document}
\title{Tag-Cloud Drawing: Algorithms for Cloud Visualization}

\numberofauthors{2}

\author{
%
\alignauthor Owen Kaser\\
       \affaddr{University of New Brunswick}\\
       \affaddr{Saint John, NB, Canada}\\
       \email{o.kaser@computer.org}
\alignauthor Daniel Lemire\\
       \affaddr{Universit\'e du Qu\'ebec \`a Montr\'eal}\\
       \affaddr{Montreal, QC, Canada}\\
       \email{lemire@acm.org}
}

\date{\today}

\newcommand{\owen}[1]{\textcolor{blue}{#1}}
\newcommand{\daniel}[1]{\textcolor{red}{#1}}
\renewcommand{\owen}[1]{}
\renewcommand{\daniel}[1]{}

\maketitle
\begin{abstract}
Tag clouds provide an aggregate of tag-usage statistics.
They are typically sent as in-line HTML to browsers.  However, display
mechanisms suited for ordinary text are not ideal for tags, because
font sizes may vary widely on a line. 
As well, the typical layout does not account for relationships that
may be known between tags.  
This paper presents models and algorithms to
improve the display of tag clouds that consist of in-line HTML,
as well as algorithms that use nested tables to achieve a more general
2-dimensional layout in which tag relationships are considered.
The first algorithms leverage prior work in typesetting and rectangle
packing, whereas the second group of algorithms leverage prior work
in Electronic Design Automation.  Experiments show our algorithms can
be efficiently implemented and perform well. 

\end{abstract}

\category{H.3.3.}{Information Storage and Retrieval}{Information Search and Retrieval}



\section{Introduction}


Tag clouds have become a popular method to support navigation 
and retrieval of tagged data.  This paper seeks to optimize the display
of tag clouds, without concern for the origin
of the tags.  We follow Hassan-Montero and
Herreno-Solana
\cite{hass:improving-tag-clouds} in assuming that associated
tags ought to be placed near one another.

Tag clouds 
conceptually resemble  
histograms, but whereas 
histograms are typically used to represent the frequencies of perhaps 
a dozen items, tag clouds can represent the 
frequencies of a hundred items. By displaying tags as hyperlinks, we
obtain interaction possibilities lacking in histograms.
In collaborative tagging, users 
are motivated to contribute tags to change the appearance of the tag
cloud~\cite{1105676}.

\begin{figure}
\subfigure[\label{fig:flickr}Flickr's  ``all time most popular tags'']{
  \fbox{\includegraphics[width=0.6\columnwidth]{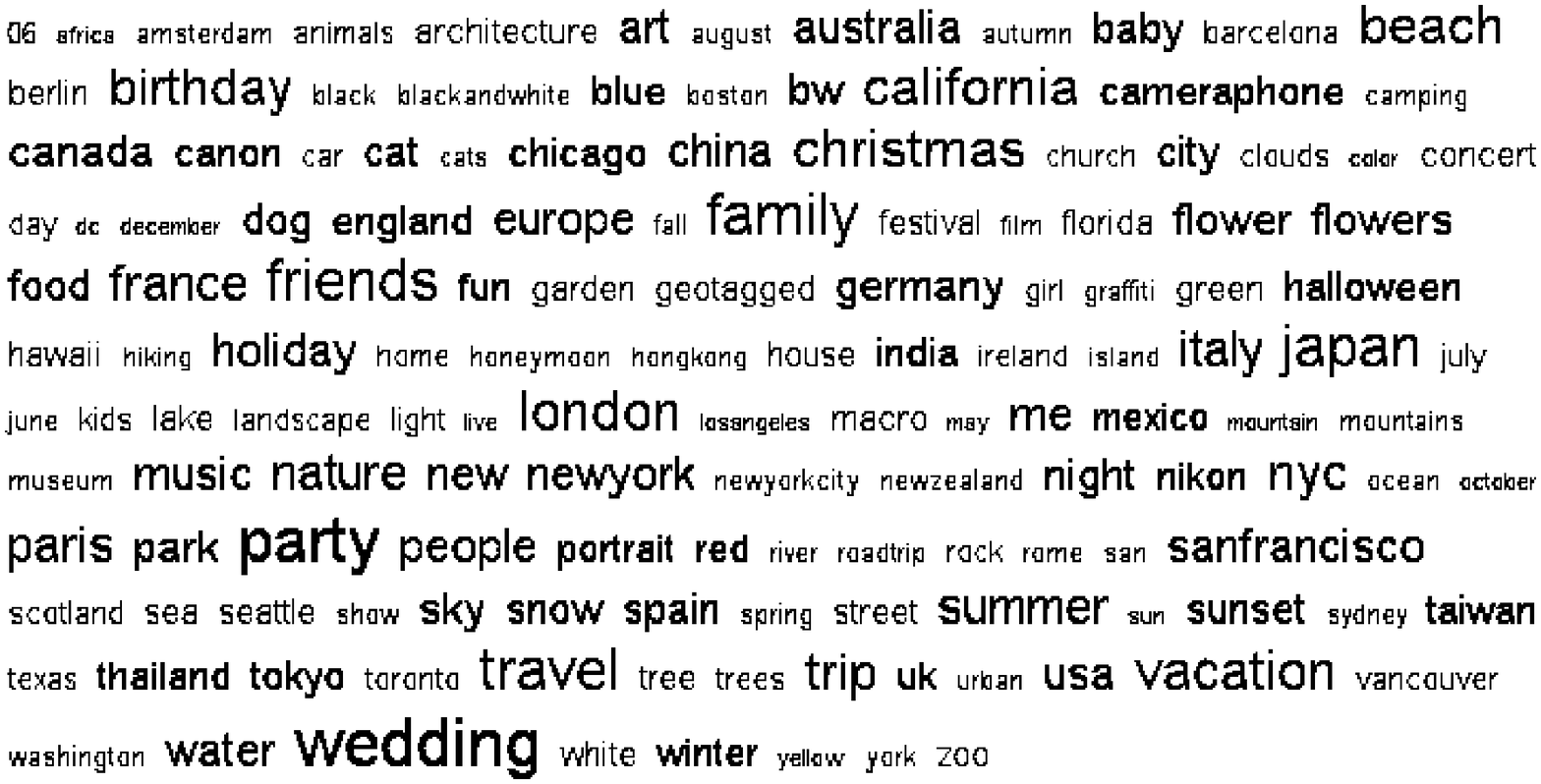}}
}
\subfigure[\label{fig:zoom}Sample from ZoomClouds.com]{ 
\includegraphics[height=0.38\columnwidth]{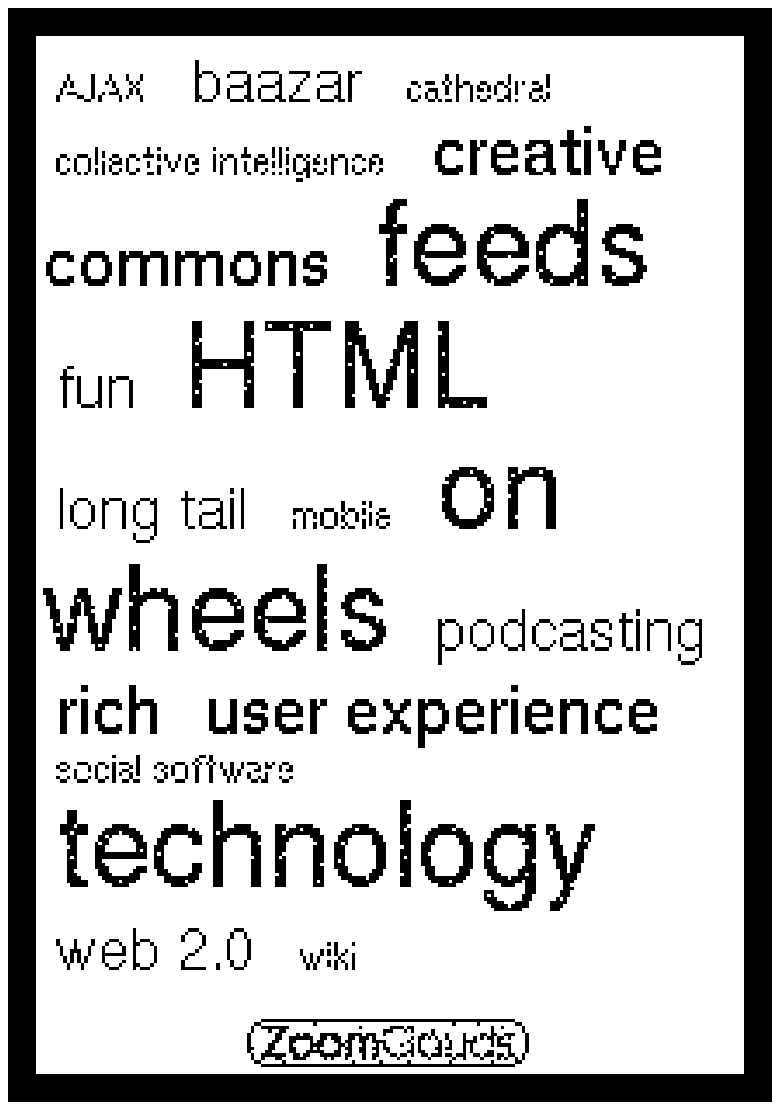}
}
\vspace{-0.2cm}

\subfigure[\label{fig:techno}Technorati's  ``Top 100 Tags'']{  \fbox{\includegraphics[width=.92\columnwidth]{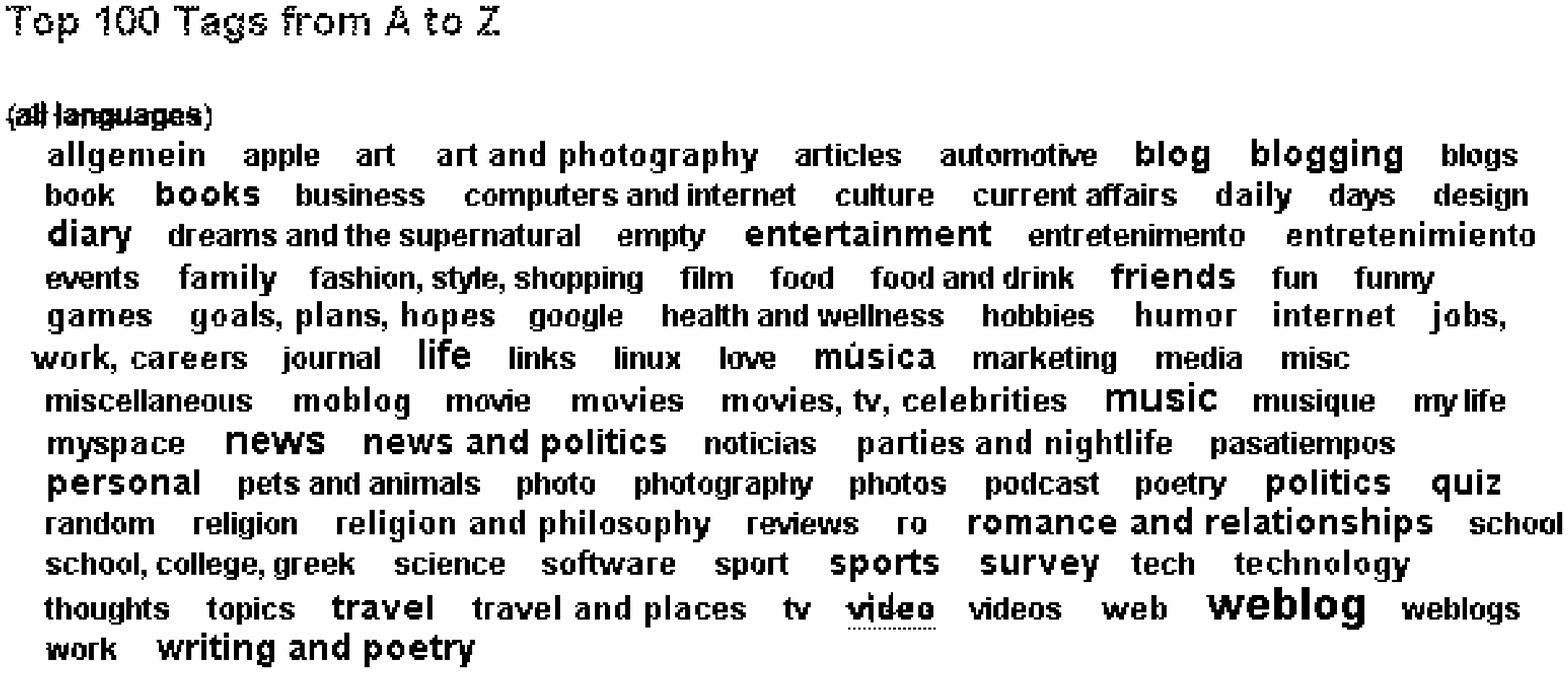}}
}
\vspace{-0.2cm}
\centering\subfigure[\label{fig:clusty}Result from Clusty.com, search ``tag cloud'']{ 
\centering\includegraphics[width=.97\columnwidth]{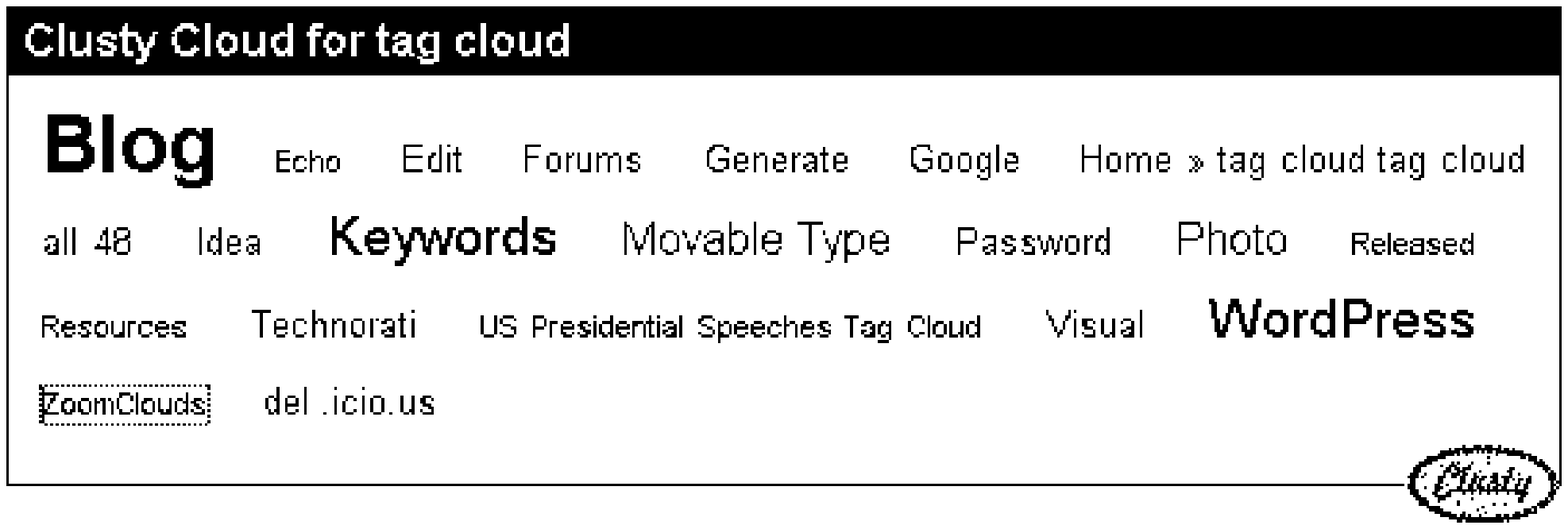} 
}

\caption{\label{fig:ugly}
Tag clouds from  popular Web sites.
}
\end{figure}

Since the font size of a displayed tag is usually used to show the
relative importance or frequency of the tag,  a typical tag cloud
contains large and small text interspersed.
A consequence is wasteful white space that is problematic in 
small-display devices, 
such as PDAs and cell phones, or in tight HTML designs.
Moreover, clumps of white space 
are not, in the authors' opinion, \ae{}sthetically pleasing
(see Fig.~\ref{fig:ugly}).  
As users, we prefer to work with attractive information displays, but
inline text support in HTML is designed for paragraphs, not clouds.
While prettier tag clouds can be generated using images, 
browser-specific technologies (ActiveX),
plugins (Flash), or complex HTML (using absolute positioning),
using only simple HTML with inline text or tables remains the
commonplace approach and offers interesting challenges.

This paper tackles the two problems of wasted space and large clumps of
white space identified above.  It also considers 
grouping related tags. 
We solve the first problem
by noting that it is essentially the same as the
floorplanning/placement problem that has been tackled for at least
30 years in the field of electronic design automation (EDA).  Thus, we
propose the classic EDA algorithm, 
\emph{min-cut placement}~\cite{brue:mincut-placement},
for area minimization and clustering in tag clouds.  The result
looks unconventional because tags are not placed on lines,
but it is supported by nested HTML tables.

The second problem's solution has a conventional appearance that
does not disrupt the left-to-right, top-to-bottom order of tags.
It is a hybrid of the classic Knuth-Plass 
algorithm~\cite{knut:breaking-paragraphs}
for text justification, and a book-placement exercise
considered by Skiena~\cite{skie:algs-design-manual}.  
In the authors' view, the resulting tag clouds are visually improved
and tighter.

\section{Related Work}

Tag clouds have been attributed~\cite{wiki:tag-cloud} to
Coupland~\cite{coupland1996m} but have been popularized
by the Web site Flickr~\cite{flickr} launched in 2004. They have
since appeared on  numerous Web
sites including Technorati~\cite{technorati},
del.icio.us~\cite{del.icio.us}, and so on. While a mere visual
representation technique, tag clouds are commonly associated with
folksonomies and social software.

Graph drawing~\cite{dibattista1999gda} is a branch of graph theory
typically concerned with the generation of two-dimensional
representations of graphs that are easy to understand and pleasing to
the eye.  While there is no absolute metric for \ae{}sthetic,
experimental evidence suggests it is important to minimize the number
of edge crossing~\cite{purchase2000eiv}. Other metrics include
symmetry, orthogonality, maximization of minimum angles, and so on.

Unlike graph drawing, tag-cloud drawing has received little
attention. Hassan-Montero and
Herrero-Solana~\cite{hass:improving-tag-clouds} have proposed
improving tag-cloud layouts by clustering similar tags together and
discarding some tags. Millen et al.~\cite{1124792} have proposed that
the user be dynamically able to remove the less significant tags; they
have also added an index so that tags can found faster in large
clouds.  Bielenberg~\cite{Bielenberg2005} has proposed circular
clouds, as opposed to the typical rectangular layout, where the most
heavily weighted tags appear closer to the center.  However, clouds
are only one specific instance of tag representation.  For example,
Dubinko et al.~\cite{dubi:tags-over-time} have proposed a model to
represent tags over a time line whereas Russel~\cite{1141859} has
proposed \texttt{cloudalicious}, a tool to study the
evolution of the tag cloud over time. Jaffe et al.~\cite{1178692} have
integrated tag clouds inside maps for displaying tags having
geographical information, such as pictures taken at a given
location. 

The problem of improving the layout of HTML pages through
special-purpose algorithms has received some attention: Hurst et
al.~\cite{hurst:continuous} 
showed that it is
possible to make HTML tables significantly more appealing.
Generally, there is ongoing work to improve the layout of text in HTML
pages using Cascading Style Sheet (CSS)~\cite{css3-text}.

\section{Background}

The current paper builds on previous work in automated typesetting and previous
work in EDA.  Minimal EDA background is required
 to appreciate our claim that tag-cloud layout
can be accomplished with EDA tools, 
but more is required  for a deeper
understanding of the details and limitations of our approach.

\subsection{Typesetting}
\label{sec:typesetting}
Automatic typesetting systems, such as \TeX~\cite{knut:tex}, must
quickly fit text onto the page.  The result must be visually
attractive.  We should break lines so that  
there is an even amount of space between words.

A greedy approach fits as many words per line as possible, beginning a
new line whenever further words cannot be placed on the current line,
with the possibility of sometimes slightly squeezing the spaces between
words and letters or  hyphenating a word. 
Sneep~\cite{sneep2005} reports that this is the approach used
by most Web browsers (Microsoft Internet Explorer, Firefox, Apple's
Safari, and Opera) and most word processors. We know of no Web browser
that can hyphenate text.  Our own investigations of the Firefox~2.0
browser lead us to believe that, for English text, line breaking is
achieved using a simple greedy approach, since no squeezing of spaces
between words or letters was observed, and text justification of a
sequence of inline elements is achieved by inserting unreported pixels
between some elements.  An advantage of the greedy approach is it can
be done on-line, without waiting for the end of a paragraph.  Indeed,
a browser should start displaying content before the page has been
completely loaded.  Unfortunately, the greedy approach can (and
frequently does) produce suboptimal solutions.

For \TeX, Knuth and Plass~\cite{knut:breaking-paragraphs} compute an
optimal solution elegantly, using dynamic programming. Given a
line of text, the difference between the preferred width of the text
as dictated by the chosen font and the page (or column) width is used
to compute the \emph{badness} of the fit.
Additional penalties handle hyphenated words and
variations in the tightness of lines.  Their \textit{total-fit}
algorithm minimizes the sum of the squares of  each line's
badness.  Excluding hyphenation and penalties, we summarize their
algorithm. We label the words of a paragraph from 1 to $n$.
Let $b_{k,j}$ be the badness measure resulting from a line
containing the words $k$ to $j$ inclusively
with the convention that $b_{k,j}=0$ when $k>j$. 
Let $t_j$ be the minimal possible sum of squares of the line
badnesses when the $j^\textrm{th}$
word ends a line with the convention that $t_0=0$. We have that  $t_j = \min_{k\leq j} (
t_k + b_{k+1,j}^2)$ with an exception if $j=n$: the last line
can be shorter without the same type of penalty. For
$j>1$,  \owen{$j$, not $i$, right?}
let $K_j=\arg
\min_k ( t_k + b_{k+1,j}^2) $ be the last word of the  line
prior to the one ending with the $j^\textrm{th}$ word. We can compute
$K_j$ for all possible $j=1,\ldots, n$ in time $O(n^2)$ and $O(n)$ space.
  We can then  reconstruct the
optimal solution recursively with the following line breaks: 
$n,
K_{n}, K_{K_{n}},\ldots,0$.

If our tags must be presented in a given order, and if all tags
have the same height, then this approach can be used to lay out a 
cloud optimally.  However, 
clouds have tags
of various heights which can be reordered, colored, etc.

\subsection{EDA: Physical Design}

Techniques for electronic design automation (EDA) have received much
research attention in the past few decades.  Within the EDA field,
\emph{physical design} of VLSI refers to the process of translating
from high-level logical circuit descriptions down to a specification
of the locations and shapes of individual transistors, wires, and so
forth.
Today, designs are frequently composed of a
mixture of custom-designed blocks of circuitry and licensed ``Intellectual
Property (IP) blocks'' 
of pre-designed circuitry.  See
Lengauer~\cite{leng:combinatorial} for more information on physical
design.   

\emph{Placement} and \emph{floorplanning} are two closely related
stages during many physical design flows.  Both concern the assignment
of blocks of circuitry to locations on the chip.  For instance, two
submodules in a design might include (rectangular) IP-blocks for a ROM and
a shift register.  Placement/floorplanning might decide that the ROM
should have its lower-left corner at (0,0) on the chip, rotated by 90
degrees, and the shift register should be rotated 180 degrees and have
its lower-left corner at (200,200).  This decision avoids
module overlap and leaves enough blank space for
the interconnection wires between them that the subsequent
\emph{routing} phase can succeed, while not leaving
 excessive space between the items, as small chips are
preferred.

While it is sometimes observed~\cite{DBLP:journals/tcad/RoyAPM06}
that, mathematically, floorplanning and
placement solve the same problem, from a practical viewpoint they are
applied differently.  Floorplanning is often done early in the design
stage, sometimes before the designs of the submodules are begun.
Using estimates of the area required for submodules (and constraints
on the aspect ratio), during floorplanning we not only choose module
locations, but we also choose module shapes.  Then the modules can be
custom designed according to the required shapes.  As module design
progresses, more accurate shape estimates may require that
floorplanning be re-done.  
Floorplanning gives a
``bird's eye'' view of the layout, based on incomplete area and wiring 
estimates.  Placement, on the other hand, is typically done with complete
knowledge of module shapes, the locations of interconnect ``pins'' on
the boundaries of the modules, and so forth.

The scenario presented assumed that floorplanning is done with
\emph{soft modules} whose aspect ratios can vary as needed.
IP blocks give rise to \emph{hard modules}, whose shape
cannot be adjusted.  A further case arises in floorplanning when a
collection of logically equivalent hard modules are available.

A final distinction between floorplanning and (final) layout is that the
former is iterated, often while a human designer is exploring 
design alternatives.  Thus, floorplanning must be fast.   In 
contrast, during final placement, the quality of solution is more
important than the running time.

Despite the conventional distinction of floorplanning from placement, 
recent tools~\cite{DBLP:journals/tcad/RoyAPM06} blur the distinction.

\subsubsection{Placement Approaches in EDA}

Placement problems are typically NP-hard: even 2-d packing
problems that ignore routing are intractable~\cite{lodi2002tdp}.
Therefore many heuristics have been proposed.  Approaches
include force-directed placement (e.g., 
considered recently by Kennings
and Vorwerk~\cite{kenn:force-directed}), where modules are 
 attracted to modules with which they are strongly
interconnected, and repulsed by nearby modules in general (to try
to reduce overlap). 
Force-directed methods have been adapted for  graph 
drawing~\cite{eade:force,fruc:force-graph-drawing}.
When solution quality is more
important than speed, metaheuristics such as simulated
annealing~\cite{kirk:simulated-annealing-science} 
are often used to guide
semi-exhaustive searches.  Such approaches would be justified for
clouds computed once and accessed many times.  Consider a
tagging site's list of hot tags for the previous month, optimized for
a common display size.

For speed, \emph{min-cut
placement}~\cite{brue:mincut-placement} is often chosen.  Since we
envision tag clouds generated on-the-fly by a server, we adapt min-cut
placement to tag-cloud display.

\section{Models for Cloud Optimization}

We consider two \ae{}sthetic models, one for
tags as  inline text and one for tags in nested HTML table.

\subsection{Tag Clouds with Inline Text}
\label{model-inline}

A tag cloud with inline text is a paragraph (block) made exclusively
of inline HTML elements such as \texttt{span}, \texttt{font},
\texttt{em}, \texttt{b}, \texttt{i}, \texttt{strong}, \texttt{a}, and
\texttt{br}. A tag, even one with spaces, must remain on a single
line.  White space outside the tags is in a given default font and
font size. Any area outside a tag, but inside the tag cloud will be
referred to as ``white'', irrespective of the background color.
  The fonts and font sizes corresponding to different tags
are enforced using inline elements with, for example, the HTML
\texttt{style} attribute.  The width available to the tag cloud is
also determined depending on the page layout, but the height of the
tag cloud is assumed to be a free parameter. Naturally, the fonts and
font sizes as well as the tag-cloud width are determined by the Web
browser as well as by the page content.  While the CSS properties
\texttt{letter-spacing} and \texttt{word-spacing} allow us to change the width
of phrases, there are implementation-specific limitations.  Our
primary view has the width and height of each tag fixed, although
we consider relaxing this restriction in Sect.~\ref{sizing}.
Similarly, the horizontal space between tags must be at least
as large as the normal space in the default font. Hence,
we will not include a penalty for squeezing tags or spaces. This is
in line with the current breed of Web-browser layout engines.

While tags are commonly ordered  alphabetically in 
clouds, we find no evidence that
users actually browse tag clouds alphabetically. For large
clouds,  a simple ECMAScript search box highlighting tags
starting with some text can make searching
specific tags convenient~\cite{1124792}.
\owen{preceding sentence could be omitted since this is mentioned
in related work.  However, the resulting paragraph is too short.}

Let the height and width (in pixels) of the $k$ tags on some line be  $w_i,h_i$ for
 $i$ ranging from 1 to $k$.
The height $h$ of the line is determined by the tallest tag in the line
($h= \max h_i$)
whereas $w$, the width of the cloud, is fixed. 
For each line of the tag cloud, there might be extra horizontal white space 
$\omega =w-\sum w_i- (k-1) W$ where $W$ is the normal width of a white space.
Hence, there is at least $h \times \omega$ extra white-space area on a line.
Because we fix $w$
but not the maximal width of a 
tag, 
we must permit $\omega$ to be negative 
(but only when a very wide tag is alone on a line).
Similarly, lines in text are typically separated by some white space
(dictated by the \texttt{line-height} property in CSS), but it does not enter
into our model. However,
when a tag is shorter than the tallest
tag on its line ($h_i < h$), this introduces some (extra) vertical space above the tag having area $(h-h_i)w_i$. Therefore,
in our model  (see Example~\ref{modelexample}), we define the badness
of a line as $h \times \vert \omega \vert + \sum_i (h-h_i)w_i$ where the sum is over the tags on 
the line. Hence, the badness of a line is only a function of the set of tag
dimensions ($w_i,h_i$).

This badness measure does not take into account symmetry or
homogeneity. In fact, the exact placement of the tags on the line is
not measured: tags can be left aligned, centered or justified. Lines
can be permuted without changing the badness. The alignment of tags
across lines, as a measure of orthogonality, is also not taken into
account.  Finally, for clouds with inline text, the order of text is
presumed either fixed or unimportant.

\begin{example}\label{modelexample}
Suppose that the tags on a line have the following 
sizes in (width, height) format: (32,14), (45,16), (24,12)
with a specified tag cloud width $w$ of 128~pixels
and an expected white-space width of 4~pixels between tags.
The line height is $h= \max\{14,16,12\}=16$. There
is extra (horizontal) white space on the line, $128-2\times 4 - 32
-45-24=19$, contributing to the badness by $19\times 16=304$.
The first and last tags have lesser heights than the second tag,
and they contribute
respectively $32(16-14)=64$ and $24(16-12)=96$ to
the badness. The total line badness is thus $304+64+96=464$.
As another example, if we have a single tag with dimension (130,16),
then the (overfull) line has badness $16(130-128)=32$.
\end{example}

In the spirit of the Knuth-Plass total-fit algorithm~\cite{knut:breaking-paragraphs}, 
we might define the overall badness of a tag cloud as the sum of the squares 
of the badnesses of each line. Merely summing the line badnesses,
without taking the squares, is also an option.
Summing the squares of the badness has the benefit of 
penalizing more heavily solutions with some very bad lines,
whereas a straight summation might tend to produce shorter
clouds.
Or we might
minimize the maximum badness across
all lines, but this might generate very tall clouds
because if even a single line is forced into
having a large badness, then all other lines
can have the same badness without prejudice to the
overall measure.
Recall that the $l_p$ norm of a vector $v$ is defined as
$\Vert v \Vert_p  = \sqrt[p]{\sum_i \vert v_i  \vert^p}$ when
$1\leq p<\infty$ and as $\max_i \vert v_i  \vert$ when $p=\infty$.
The three aggregates above can be described by the
$l_2$, $l_1$, and $l_{\infty}$ norms respectively.

\subsection{Tag Clouds with Arbitrary Placement}
\label{model-arbitrary}

Our model for this section assumes that
\begin{enumerate}\setlength{\itemsep}{0.5ex}
\setlength{\parskip}{0.5ex}
\item tags may be reordered and placed arbitrarily (but without overlap or rotation) in the plane;
\item tag relationships are known, and strongly related tags should be in close
proximity;
\item tag-cloud width has an upper bound;
\item tag-cloud height should be small, to reduce scrolling;
\item (optional) tags may be deformed slightly (made shorter but wider,
for instance), so long as tag area remains (nearly) constant;
\item (optional) large clumps of white space are bad.
\end{enumerate}

In contrast to clouds with only inline text, there is no analogue to
a ``line'' of tags when arbitrary placement is allowed.
  Therefore, when adapting the model
from Sect.~\ref{model-inline}, it is not clear how to combine the
various undesirable white  areas that are
in excess of a tag and its small surrounding border.
A simple and appealing method is to sum this bad area, which is 
equivalent to minimizing the area occupied by the tag cloud.

Another (possibly conflicting) goal is to obtain spatial clustering of
semantically related tags. If we form a graph with tags as vertices and
edge weights indicating the strength with which two tags are linked, a
reasonable measure of (undesirable) spatial non-proximity is
\begin{equation}
\label{wgtd-wirelen}
\sum_{p,q} w(p,q) d(p,q),
\end{equation} where $p$ and $q$ are placed tags linked
with strength $w(p,q)$ and separated spatially by distance $d(p,q)$.
Small values indicate better clustering.  In experiments, we used
Euclidean distance for computing $d(p,q)$.

\subsection{Tag Relationships}

The previous subsection assumed a graph-based model with a binary
tag-similarity relation.  
Yet, higher-degree relations may also make sense, loading
to a hypergraph-based model.  

One method of determining tag relationships counts
co-occurrences~\cite{hass:improving-tag-clouds}, when a pair of
tags have been assigned to the same resource (e.g., a photo).  Viewed this
way, relationships are binary and can be modelled as edges
in a graph.  Another view is that each resource corresponds to  a hyperedge
in a hypergraph, whose members consist of the tags.
Translation between these views can be achieved by replacing each 
hyperedge by a clique.

For example, 
consider
a resource (perhaps a photo) tagged ``baby, tears,
bottle, diaper'', another tagged ``bottle, gas, beer'', another tagged
``beer, rioting, sports'', and a fourth  tagged ``gas, tear gas,
tears, rioting''. (See Fig.~\ref{bipartitioningA}.)
The second view has 4 hyperedges, whereas the first view
has 
${4 \choose 2} + { 4 \choose 2} + {3 \choose 2} + {3 \choose 2}$~edges.
  For instance, the hyperedge
\{bottle, gas, beer\} from the first view would correspond to the
edges \{ (bottle, gas), (bottle, beer),  (gas, beer) \} in the second
view.

In EDA, the r\^ole of tags is played by modules and the natural
relation between modules is ``have a wire that interconnects several
modules'',
leading to a hypergraph model.  We argue in Sect.~\ref{placementdiffs}
that tag-cloud display should instead use a graph model.

\section{Solutions}

We propose different approaches to the two major problems.  For inline
text, dynamic programming or shelf-packing heuristics can be applied.
For arbitrary placement, we use the min-cut placement algorithm from EDA.

\subsection{Cloud Layout with Inline Text}

Our first breed of algorithms 
take an ordered list of tags and choose where to break lines.
We first designed a simple greedy algorithm: tags are added
to the current line one by one, inserting a white space between
them, until the line is full. It runs in $O(n)$ time
and matches
what is done by most browsers. When a tag is too wide to fit on even 
an empty line,
a new line is created for this tag alone.
Second, we implemented a dynamic-programming solution.
Our algorithm is nearly identical to the $O(n^2)$ time and $O(n)$ space
 Knuth-Plass 
algorithm~\cite{knut:breaking-paragraphs} given in Sect.~\ref{sec:typesetting},
except that:
\begin{itemize}
\setlength{\itemsep}{0ex}
\setlength{\parskip}{0ex}
 \item the last line is not an exception: it cannot be half empty without penalty;
 \item if, and only if, a tag exceeds the maximal width, then it will be given a line of its own;
no other overfull lines are allowed.
\end{itemize}

The second breed of algorithms reorders tags, attempting to decrease the badness.
Finding an optimal ordering is NP-hard: when the required horizontal white space between
tags is zero, we have the NP-hard Strip Packing Problem (SPP)~\cite{lodi2002tdp}.
As a rough heuristic to assess the influence of order, 
we  randomly shuffle tags several times 
(10 in our experiments), apply the dynamic-programming algorithm to 
place the tags optimally, and keep only the best solution.
Other simple heuristics are based on approximation algorithms 
for SPP, although SPP is only a special case of our problem.  We use
\textsc{Next Fit Decreasing Height} (NFDH) and \textsc{First Fit Decreasing Height} 
(FFDH) from Coffman et al.~\cite{coff:level-oriented}. They are 
SPP 2-optimal  and SPP 17/10-optimal, respectively,
and they run in $O(n \log n)$ time~\cite{lodi2002tdp}.
Both first sort tags by non-increasing height.
NFDH is then  the application
of the simple greedy algorithm described above. FFDH  places each
new tag on the first available line, starting from the first line ever 
created, and creating a new line at the 
end whenever necessary. 
Tags exceeding the maximal width are placed on a line of their own.
Because we typically have several tags with the same height, but different width,
we further refined FFDH to our \textsc{First Fit Decreasing Height, Weight} 
heuristic (FFDHW).  With FFDHW, tags continue to be primarily sorted by 
(non-increasing) height, but ties are broken by (non-increasing) width.

We could better assess the heuristics if we could obtain
optimal solutions to this NP-hard reordering problem. However, for interesting
clouds (e.g., $n=100$), the ordering search space is 
huge: $n!=100!\approx 9.33\times 10^{158}$.
We suspect branch and bound, or other sophisticated
enumerative approaches, are too slow even for experimental work.

\subsection{Cloud Layout with Arbitrary Placement}

Fast  arbitrary tag placement is achieved with min-cut placement, optionally
followed by floorplan sizing.

\subsubsection{Min-cut Placement}

\begin{figure}
\subfigure[\label{bipartitioningA}Bipartitioning: before]{
  \includegraphics[width=.45\columnwidth]{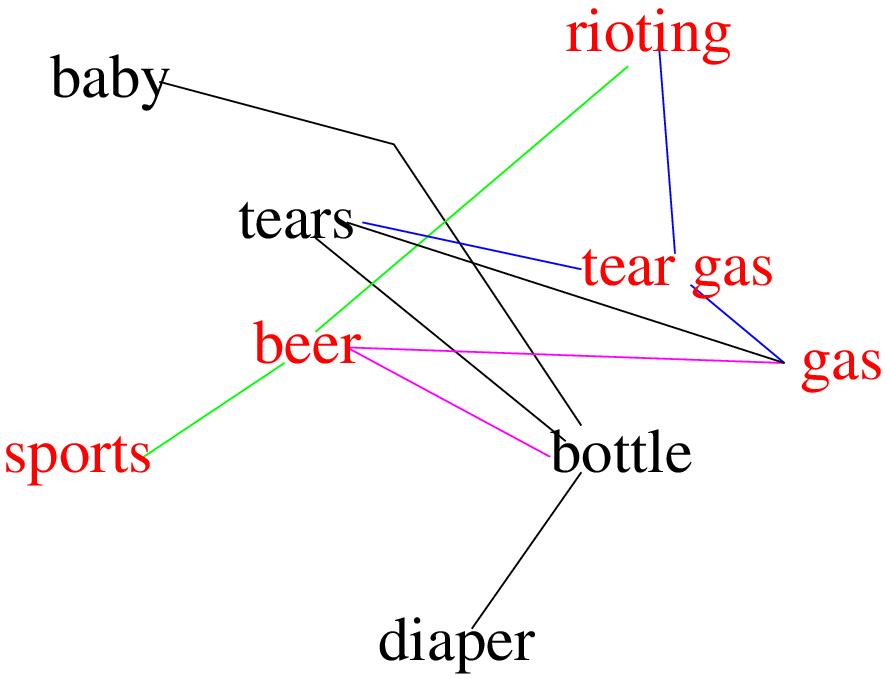}
}
\subfigure[\label{bipartitioningB}Bipartitioning: after]{
  \includegraphics[width=.45\columnwidth]{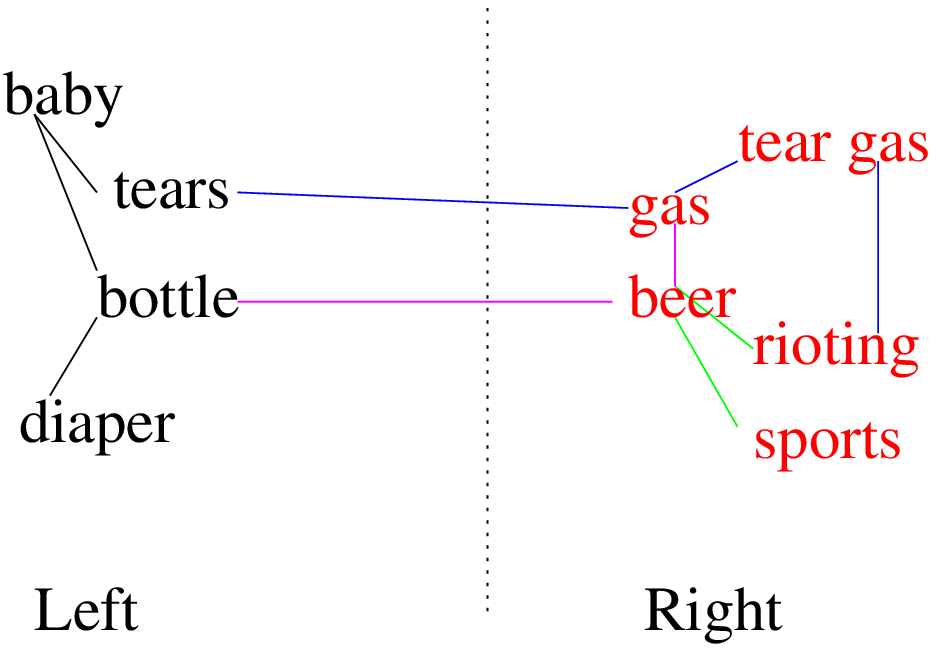}
}
\caption{\label{bipartitioning} Bipartitioning (hypergraph view). 
This cut includes two hyperedges, or
five edges. }
\end{figure}

Min-cut placement~\cite{brue:mincut-placement} recursively decomposes
a collection of tags by \emph{bipartitioning}: splitting  the 
tags into a ``Left'' group and a
``Right'' group.  Then each group is recursively
split, probably into ``Top'' and ``Bottom'' groups, although
re-splitting into ``Left'' and ``Right'' may also occur.  Ideally,
the bipartition must be fairly balanced ---the number of tags or
total areas of tags, for instance, must be similar for the two groups.
Also, the cut size (the number --- or perhaps total weight ---
of edges/hyperedges containing tags in both groups) should be small.
Since these two goals may conflict, various approaches can be
considered.  For instance, we 
specify a balance constraint and
then try to minimize the cut. (See Fig.~\ref{bipartitioning} for an
example.)  While bipartitioning is NP-hard, well-known 
heuristics exist.

\begin{figure}
{\centering\includegraphics[width=.5\columnwidth]{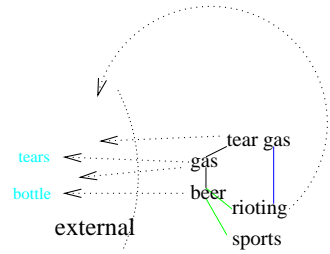}
\caption{\label{externalpulls}In future horizontal partitioning
there will be a bias to encourage all tags (except \texttt{sports})
to the left side of their area.}
}
\end{figure}

During partitioning of a group of tags, there should be an
influence of ``outside'' tags.  Therefore, we track how strongly
each tag is connected to external tags known to be above, below,
leftward, and rightward of the group of tags being bipartitioned.
See Fig.~\ref{externalpulls}, where we see that even 
though tags \texttt{beer} and \texttt{sports} are connected,
tag \texttt{beer} has an external leftward pull but \texttt{sports}
does not.   This may encourage partitions where these two tags are
separated.  Integrating external pulls into bipartitioning is often handled in
min-cut placement by creating two dummy tags, ``externalLeft'' and
``externalRight'' and insisting that these dummy tags cannot change
location~\cite{dunl:placement}.  
Tag ``externalLeft'' is a surrogate for all
external tags to the left of the nodes being partitioned.
``ExternalTop'' and ``externalBottom'' are similar.

Under reasonable assumptions about tag sizes and bipartition balance
requirements, given $n$ tags with $m \in \Omega(n)$ relationships,
min-cut placement can run in O($m \log n$) time if we use the
Fiduccia-Mattheyses bipartitioning heuristic~\cite{fidu:FM}.

\subsubsection{Slicing Floorplans}

\begin{figure}
\subfigure[\label{slicingtree}Slicing tree. Numbers next to nodes relate
to areas in the slicing floorplan.]{
  \includegraphics[width=.45\columnwidth]{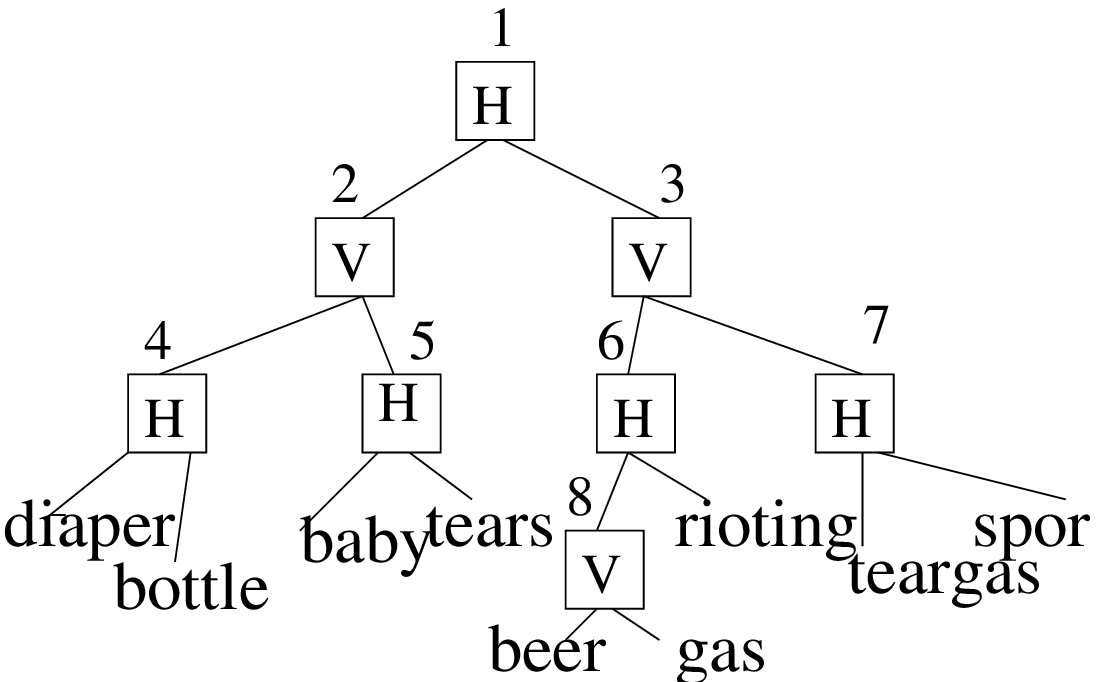}
}
\subfigure[\label{slicingfloorplan}Slicing floorplan]{
  \includegraphics[width=.45\columnwidth]{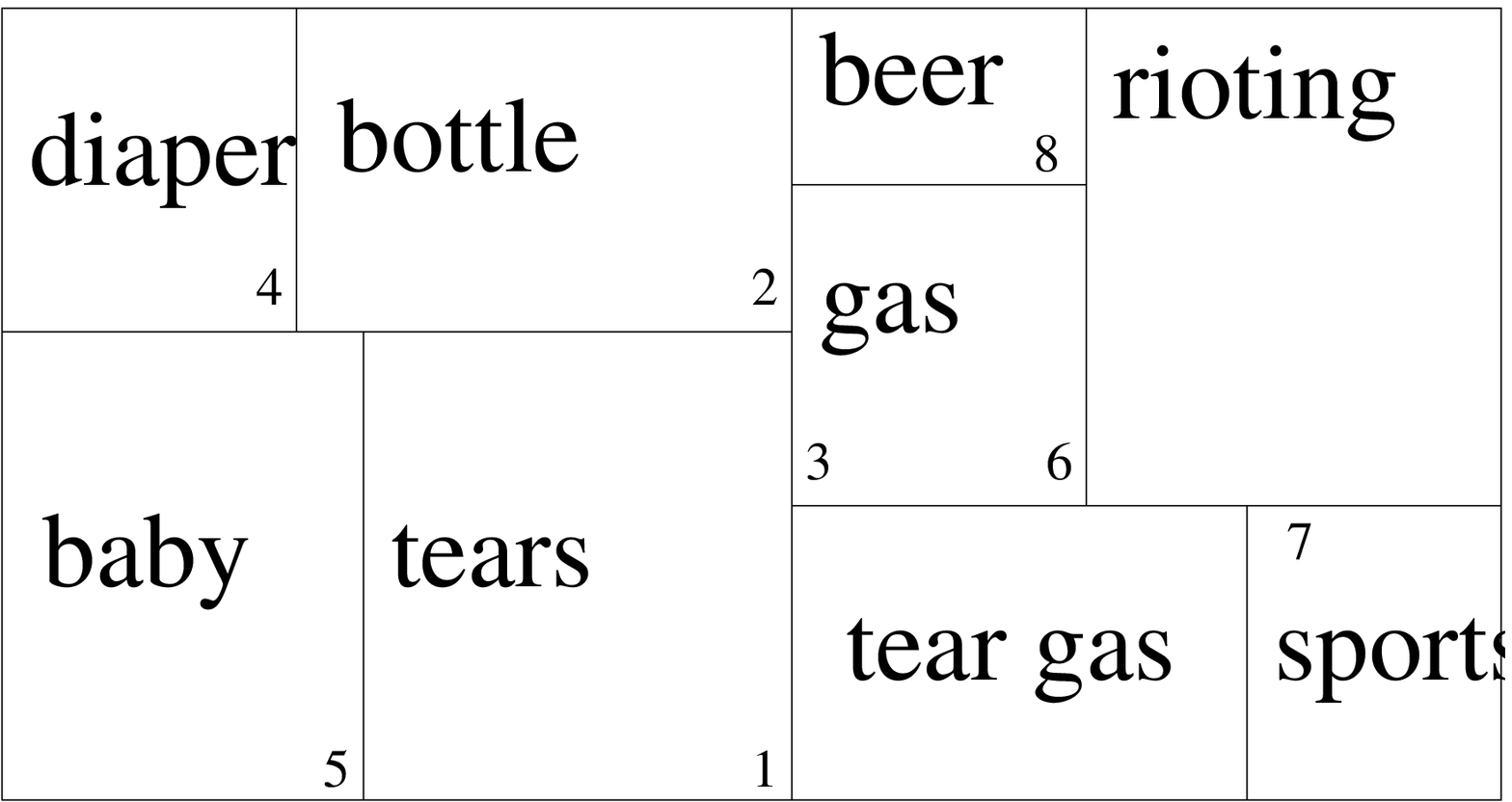}
}
\caption{\label{slicingthings} Slicing tree and  associated slicing 
floorplan.  Cut-lines numbers indicate slicing-tree nodes.}
\end{figure}

Recursive bipartitioning's effect can be represented in a
\emph{slicing tree}~\cite{stoc:orientation-floorplan}. See, 
for instance, Fig.~\ref{slicingtree}.
Leaves store tags. Internal  nodes specify
the relative placements of tags in the subtrees, and they
are labelled \textbf{H}orizontal or \textbf{V}ertical, depending 
how they divide tags.
Each internal node
is naturally associated with a placement area into which all tags in its subtree
will be stored.  The node also slices its placement area, either 
horizontally or vertically, assigning each of its subtrees to one of
the sub-areas.
At the finest level, each tag has been assigned a particular area into
which it, and only it, may be placed.  The resulting subdivision of
the placement area is a \emph{slicing floorplan} 
(see Fig.~\ref{slicingfloorplan}) and can be used for placement: 
a straightforward tree traversal
can assign precise locations to a ``tightest possible'' placement
that corresponds to the slicing floorplan. 

\subsubsection{Nested Tables for Slicing Floorplans}

\begin{figure}
\subfigure[\label{caxtons-cloud-borders}Displaying table borders]{
\centering\includegraphics[width=0.8\columnwidth]{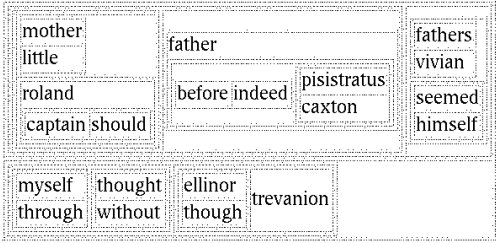}
}
\subfigure[\label{caxtons-cloud}Displayed with appropriate CSS]{
\centering\includegraphics[width=0.8\columnwidth]{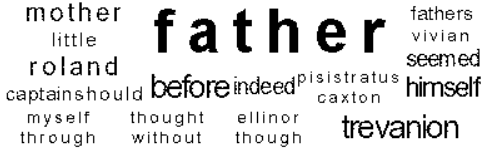}
}
\caption{\label{caxtons} Slicing floorplan shown as
nested tables.}
\end{figure}

Given a slicing tree, it is simple to make the browser
render the tag cloud. (See Fig.~\ref{caxtons}.) 
We use a trick: each internal
node in the slicing tree corresponds to a 2-element table in HTML.
The table is either $2 \times 1$ or $1 \times 2$, depending whether
the slicing-tree node is tagged `H' or `V'.  If the node's children
are not leaves, then the table's cells contain sub-tables.
For example, node~6 in Fig.~\ref{slicingtree} leads to
{  
\begin{center}
\begin{minipage}{0.6\columnwidth}
\lstset{language=HTML, emph={table}, emphstyle={\color{blue}\textbf}, 
emph={[2]tr}, emphstyle={[2]\color{red}\textbf}}
\begin{small}\begin{lstlisting}
<table><tr>
  <td> <table>
     <tr><td>beer</td></tr>
     <tr><td>gas</td></tr>
     </table></td>
  <td>rioting</td> 
</tr></table>
\end{lstlisting}\end{small}
\end{minipage}\end{center}
}

CSS (e.g., \texttt{border-spacing:0px}) then
reduces whitespace.

\subsubsection{Choosing Aspect Ratios}
\label{sizing}

Although the orientations chosen for the cuts in the slicing tree
have perhaps the largest effect on the eventual shape of the 
layout, floorplanning can also
choose \emph{which} precise shape to use.  In VLSI, there may be a
tall skinny implementation of a ROM
or a functionally equivalent
square implementation. With tag clouds, we may be willing
to stretch or squash a tag somewhat, as long as its total area remains
more-or-less constant.  This can be accomplished using 
CSS's \texttt{font-stretch}~\cite{css3-fonts}.  Unfortunately,
few browsers act on this property yet; to simulate it, we adjusted
\texttt{font-family}, 
as well as 
\texttt{letter-spacing}, \texttt{font-weight} and \texttt{font-size}.

This \emph{floorplans sizing} can be done 
efficiently~\cite{shi:optimal-slicing,stoc:orientation-floorplan}
for slicing floorplans. In particular, it runs in $\Theta(s \log s)$
time if  $s$ represents the sum, taken over  the number of shape 
options for each tag.  
Yet for general floorplans this
problem is intractable~\cite{stoc:orientation-floorplan}.

\subsubsection{EDA Placement Is Not (Quite) Tag  Placement}
\label{placementdiffs}

Overall, the EDA problem of placement/floorplanning and our problem of
tag-cloud layout are almost the same.
Can we simply feed our tag-cloud data
to an EDA placement tool and then extract a final placement?
The answer is a qualified ``yes'': we have essentially done this,
but we  found it appropriate to modify the EDA tool.

Long tags can have aspect ratios that would be unusual for
EDA.  In placement or floorplanning, it is often permissible for
cells to be rotated by 90 degrees.
With rotation, every tall, thin module can become a short, wide module
and thus there is a symmetry.  With tags, a 90 degree rotation might
be artistically interesting but hard to read.  Thus
we forbid such rotation.  Rarely do we see a tag that is taller
than it is wide;  moreover, we have constrained the cloud width (but not
height). Thus, we must handle widths and heights asymmetrically. 

In some EDA design styles, the interconnect wiring must run between
modules, rather than atop them.  Thus, adequate white space
must be allocated throughout the layout to accommodate wiring.  Superficially,
tag clouds are similar: we should not abut two tags without leaving some
white space between them.  However, in EDA the amount of white space at
any particular area depends on the number of wires that must pass
though that area.  This is much more complicated than with tags, where
a fixed boundary, or perhaps one proportional to the font size, is 
appropriate.

In both floorplans and tag clouds, strongly coupled items should be
close to one another.  For EDA, coupling comes from \emph{nets}, which
are best modelled as hyperedges in a hypergraph.  Each net is a subset
of modules, and electrical connectivity will eventually be achieved by
finding a spanning (or Steiner) tree  over the
modules belonging to the net.  The transitivity of electrical
connections is a factor: consider the wiring necessary when a single
net includes two tightly packed clusters of 10 modules each, found at
opposite ends of the layout.  A \emph{single} wire can traverse the
long distance; min-cut should count a cost of 1 for this net when
bipartitioning.  However, this behaviour is intuitively wrong when
considering one cluster of 20 related tags.  Each tag in the cluster
is related to every other tag, and thus dividing them should be much
more expensive: we are splitting a clique in an ordinary graph.

Finally, the expected input sizes and acceptable running times and
solution quality levels may be different.  Placement problems would
frequently have thousands of elements, more than any reasonable tag cloud.
Also, obtaining a high-quality solution would be more important than
obtaining a solution quickly.  On the other hand, any technique for
on-demand tag-cloud creation for servers must necessarily be fast.
``Fast floorplanners'' (such as McFarland
describes~\cite{mcfa:evaluation}) are  used interactively with only a
coarse subdivision of the design into top-level modules. These would
more closely match the input sizes and response-time requirements of
tag-cloud placement.

\section{Experimental Results}

To evaluate our methods, we obtained test data from several source,
implemented the methods, and analyzed the results they obtained.

\subsection{Test Data}

Tags and their accompanying importance levels (0-9) were obtained from
ZoomClouds and Project Gutenberg; on average, clouds had 93~tags.
For each of the 10~importance levels, we defined CSS classes
with corresponding style choices: font sizes ranged from 8\,pt to 44\,pt 
and the selected font family was arial.
However, our
techniques need the size of each tag's bounding box, and we
chose not to limit ourselves to monospace fonts.
Experience showed insufficient accuracy from predictions based on
knowing the text,  font size, and various CSS parameters. Therefore,
our programs are given tag bounding-box sizes as part
of their input.  These were obtained using ECMAScript 
and the  DOM attributes \texttt{offsetWidth} and 
\texttt{offsetHeight} applied to an  HTML \texttt{span} element.

Our requirement for browser-specific display information means that
practical use of these techniques is perhaps best done on the client
in ECMAScript,
although server-side processing (using AJAX, for instance) is not
impossible.  Our experimental program for in-line text was written in
Java, whereas the program for arbitrary placement was written in C;
thus layout times may reflect a server environment.

\subsubsection{ZoomClouds}

ZoomClouds~\cite{zoomclouds} is a Web site using the Yahoo! Content
Analysis API to produce historical tag clouds for any given RSS feed
using some content-processing heuristic.  They make available a REST
API producing an XML description of a tag cloud including tag names
and weights.  None of their tag clouds had more than 100~tags.  We
retrieved 65~different tag clouds with an average of 94~tags per
cloud. For each tag cloud, we normalized the weights with a linear
function so that they were integers between 0 and 9.  We chose most tag
clouds randomly with the random sources function of the Web site, but
we also included major Web sites such as USA Today, Slashdot, the New
York Times, L.A. Times, as well as major blogs such as Scobleizer and
Boing Boing.

\subsubsection{Project Gutenberg E-books}
Test data, including tag relationships, were also 
derived from word co-occurrences 
in 20~e-books produced by Project Gut\-en\-berg~\cite{gutenberg}.  
Initial processing removed all non\-alphabetic
characters, converted all characters to lower case, and removed
short words (those with 5 letters or less).  The remaining words became
tags.  Only 
the most frequent $k$ tags were kept (we used $k=20$, 50, 100 and 200)
in our tests. The importance $i$  of tag $T$ was determined as
$i = \lfloor 10 \frac{t-r}{f-r+1} \rfloor$, where $f$, $r$ and $t$ are
respectively the frequencies of the most frequent tag,  the least 
frequent retained tag, and the tag $T$.

Word co-occurrences determined the relationship strength between tags,
as in recent work on tag-cloud display~\cite{hass:improving-tag-clouds}.
Two consecutive words form a \emph{(distance 0) co-occurrence}. 
Each pair of tags had a relationship of strength $s$ if there 
were $s \geq 2$ such  co-occurrences in the e-book.

\subsection{Tag Clouds with In-line Text}

Our in-line text algorithms   were implemented in Java~1.5. On a
2.16\,GHz Intel Core 2~Duo processor,  we ran 5000~tests on a large
\texttt{del.icio.us} tag cloud~\cite{del.icio.us} (140~tags, presented in Fig.~\ref{fig:ugly}):
the average wall-clock running time for one tag cloud optimization was well 
under 1\,ms for all algorithms (except the 10-random-shuffle heuristic),
and under 0.2\,ms for the greedy
algorithms. Our code was not particularly optimized for speed.

\begin{figure}

\subfigure[\label{fig:greedyalphaexample}Alphabetically sorted tags, greedy algorithm]{
\includegraphics[width=0.47\columnwidth]{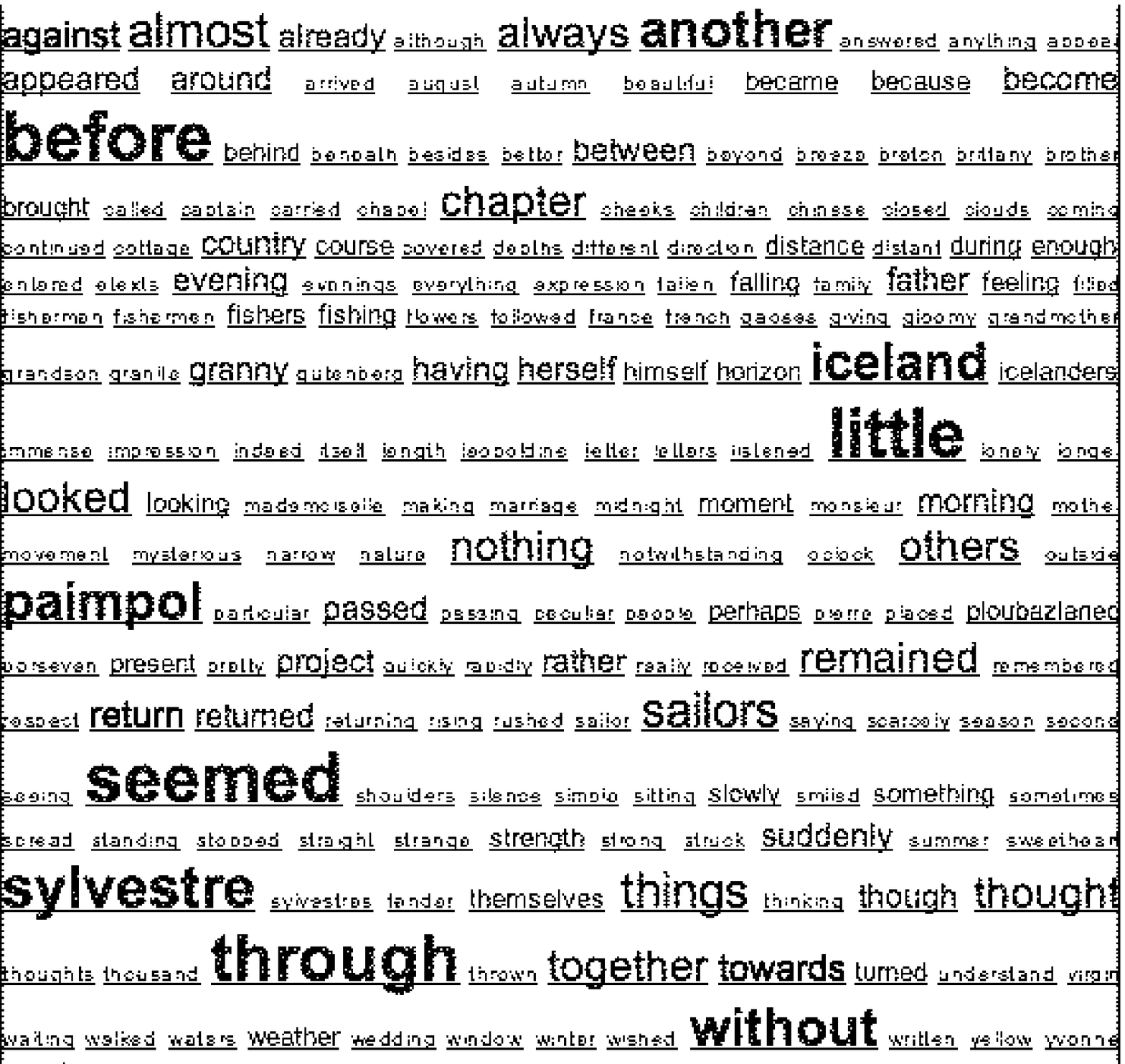}
}\subfigure[\label{fig:optialphaexample}Alphabetically sorted tags, dynamic programming]{
\includegraphics[width=0.47\columnwidth]{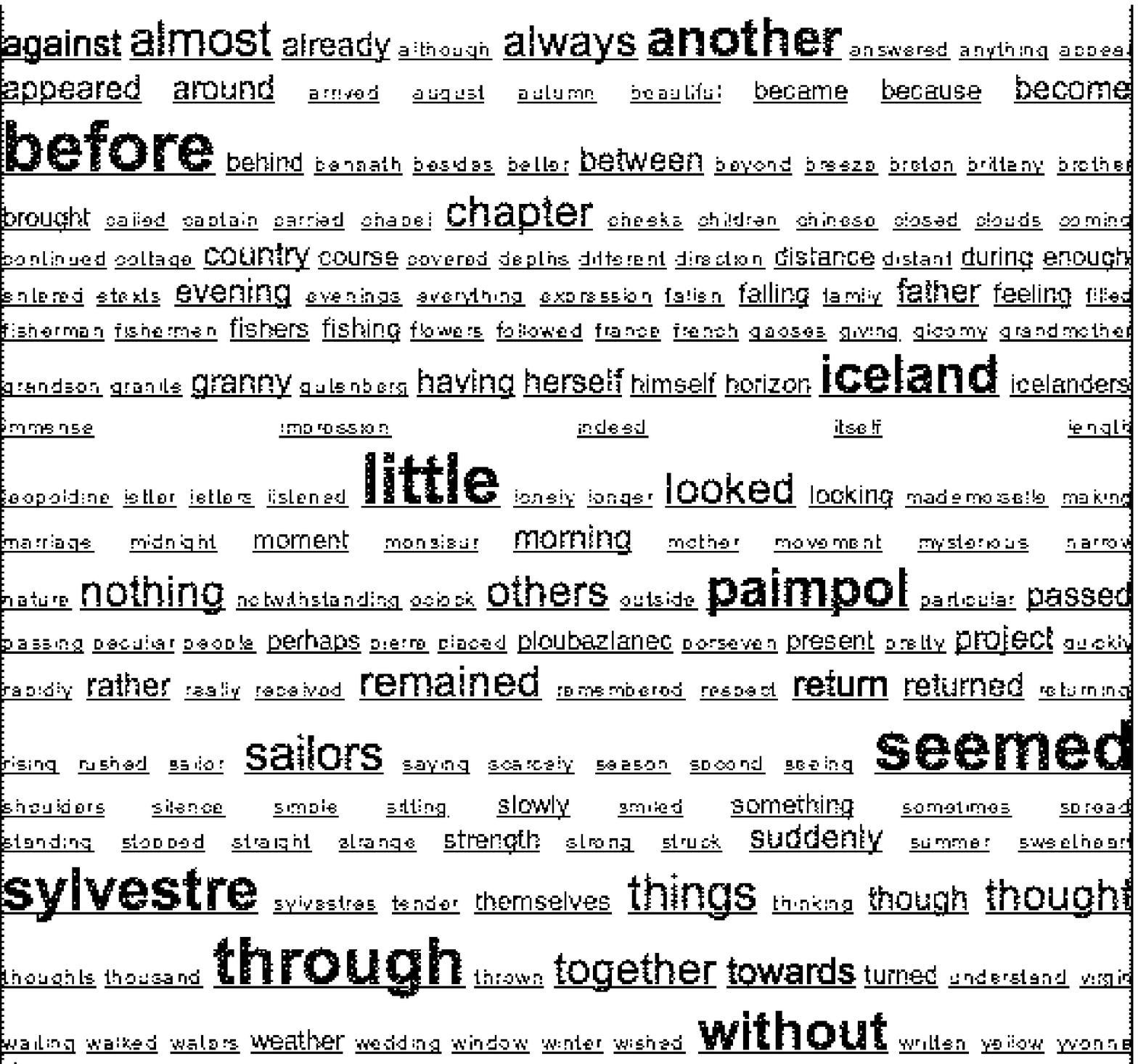}
} 

 \subfigure[\label{fig:greedyweightexample}Tags sorted by weight, greedy algorithm]{
\includegraphics[width=0.47\columnwidth]{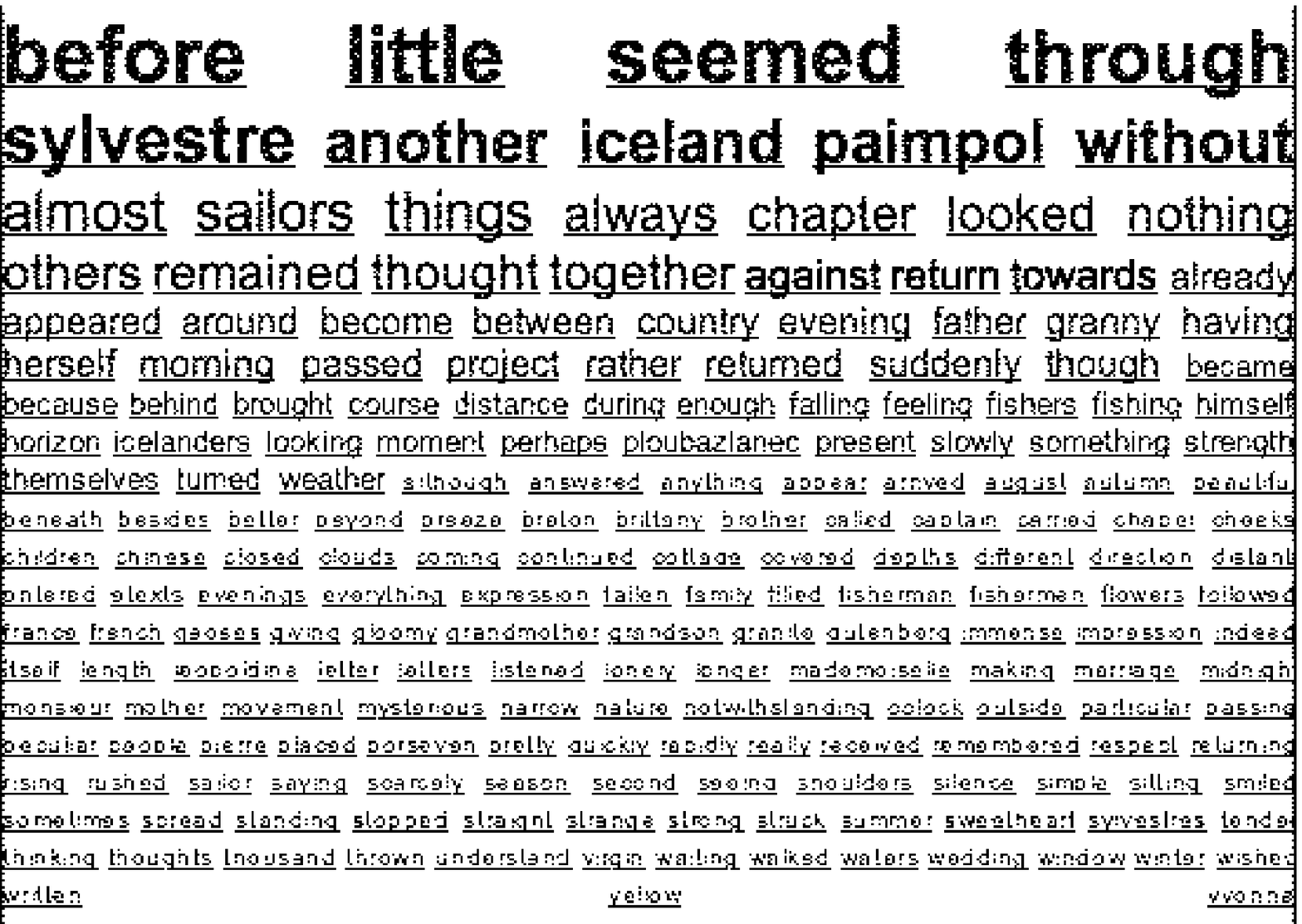}
} \subfigure[\label{fig:nfdhexample}FFDH heuristic]{
\includegraphics[width=0.47\columnwidth]{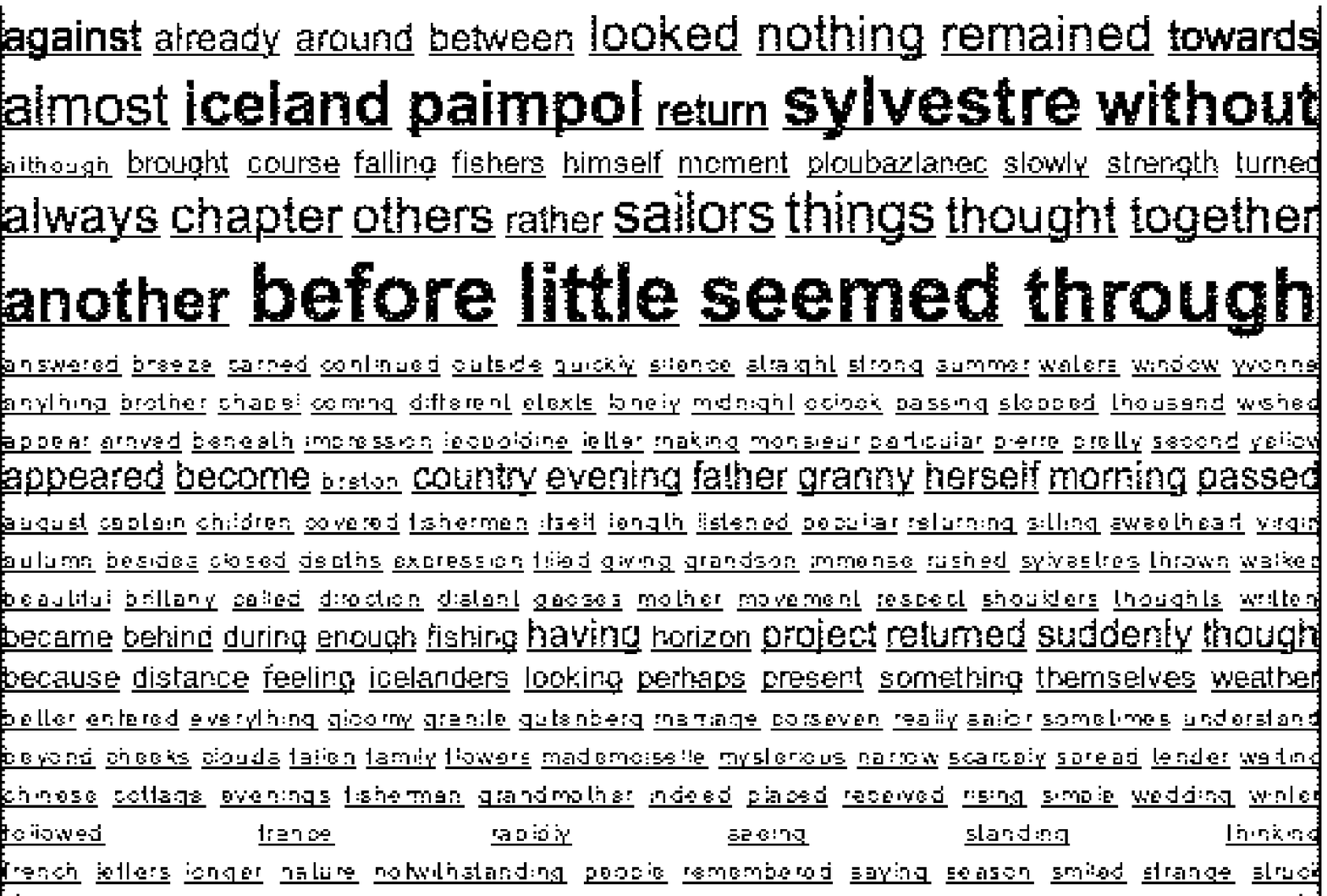}
} 
 \caption{\label{fig:algoexample}Screen shots of a tag cloud optimized using different algorithms: the greedy algorithm is similar to normal browser display.
}
\end{figure}

We  tested our algorithms  on both the 65~ZoomClouds tag clouds 
and the 80~Project Gutenberg tag clouds.
Fig.~\ref{fig:algoexample} presents a visual example of the result of
4~heuristics applied to one tag cloud. Alphabetically-sorted tags are, on average, 40\%
larger than weight-sorted tags. Dynamic programming does not
reduce the area of the tag clouds for weight-sorted tags, but offers a reduction of about 3\%
for alphabetically-sorted tags. The random-shuffling algorithm
does worse than sorting by weight\footnote{even trying 5000 shuffles (not shown).}.
The NFDH heuristic gives about the same
average tag-cloud height as does the weight-sorted greedy algorithm, but
 the FFDH and FFDHW heuristics  offer an average reduction of about 3\% in the height of
the ZoomClouds tag clouds, and of 1\% and 2\% respectively for the Project Gutenberg
tag clouds. Varying our badness model aggregates used by the dynamic-programming
algorithms shows that using the maximum line-badness aggregate
($l_\infty$) can generate unacceptably tall tag clouds (3 times taller than normal); however, 
the difference in height  between the sum ($l_1$) and sum of squares ($l_2$) aggregates is well below 1\%,
though the $l_1$ aggregate has a small edge, as expected. 
While tighter clouds do not have large clumps of white 
space, they do not necessarily appear more
symmetric.

\begin{figure}

 \subfigure[\label{fig:l1resultsgut}$l_1$ norm, Gutenberg]{
\ifpdf
\includegraphics[width=0.47\columnwidth]{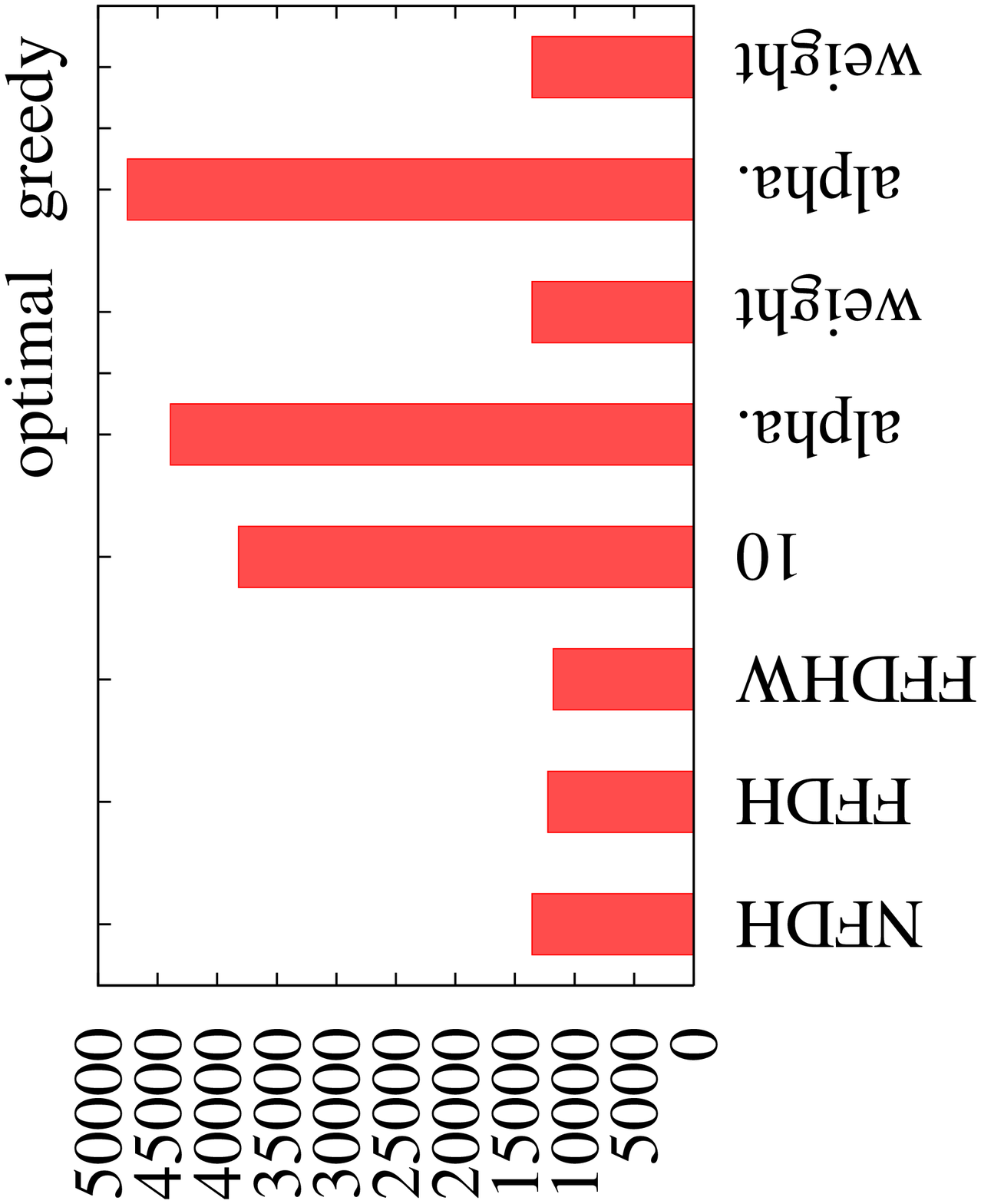}
\else
\includegraphics[height=0.47\columnwidth,angle=270]{gutclassicresults_1_0}
\fi
}\subfigure[\label{fig:l2resultsgut}$l_2$ norm,  Gutenberg]{
\ifpdf
\includegraphics[width=0.47\columnwidth]{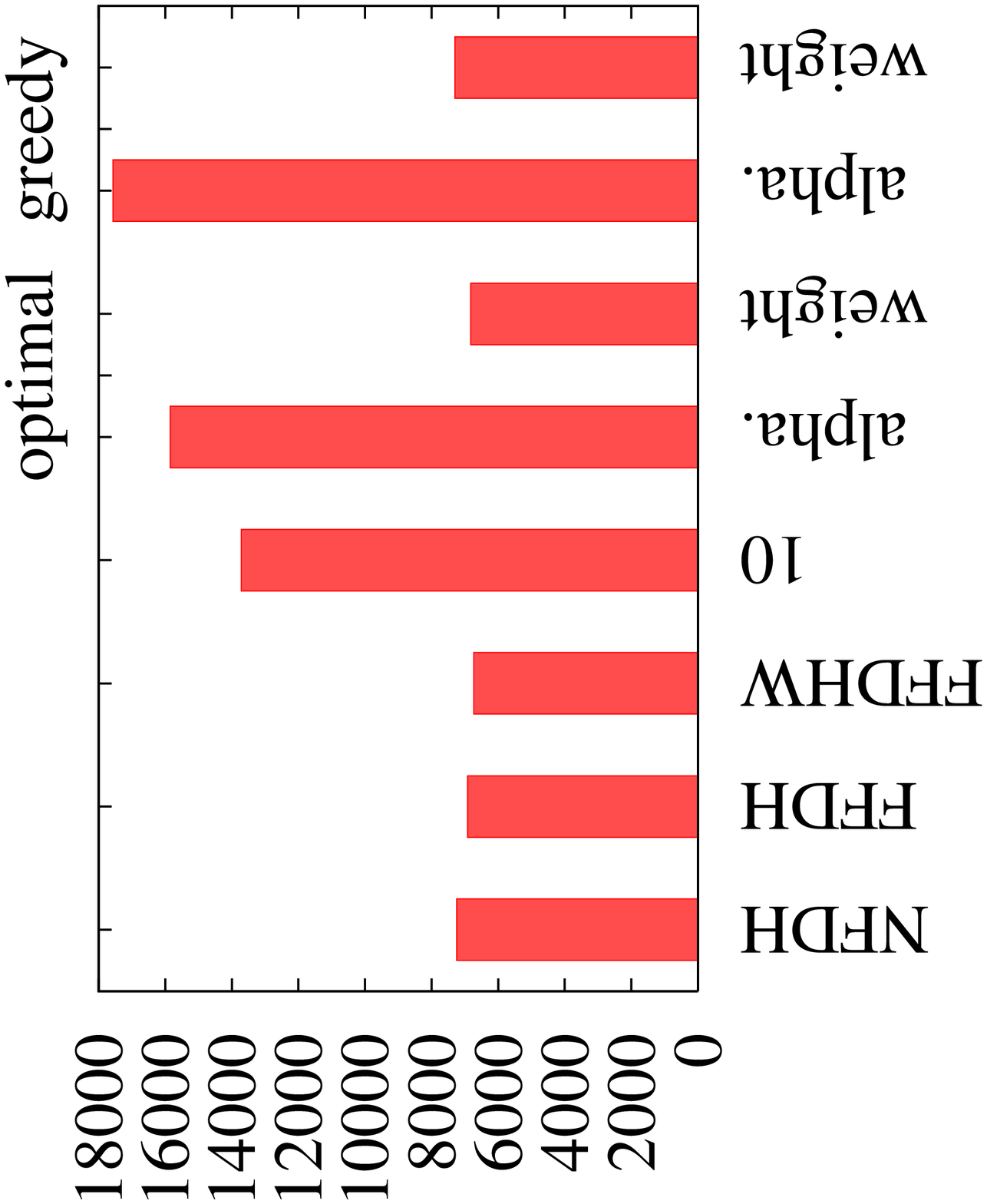}
\else
\includegraphics[height=0.47\columnwidth,angle=270]{gutclassicresults_2_0}
\fi
} 

 \subfigure[\label{fig:l1results}$l_1$ norm, ZoomClouds]{
\ifpdf
\includegraphics[width=0.47\columnwidth]{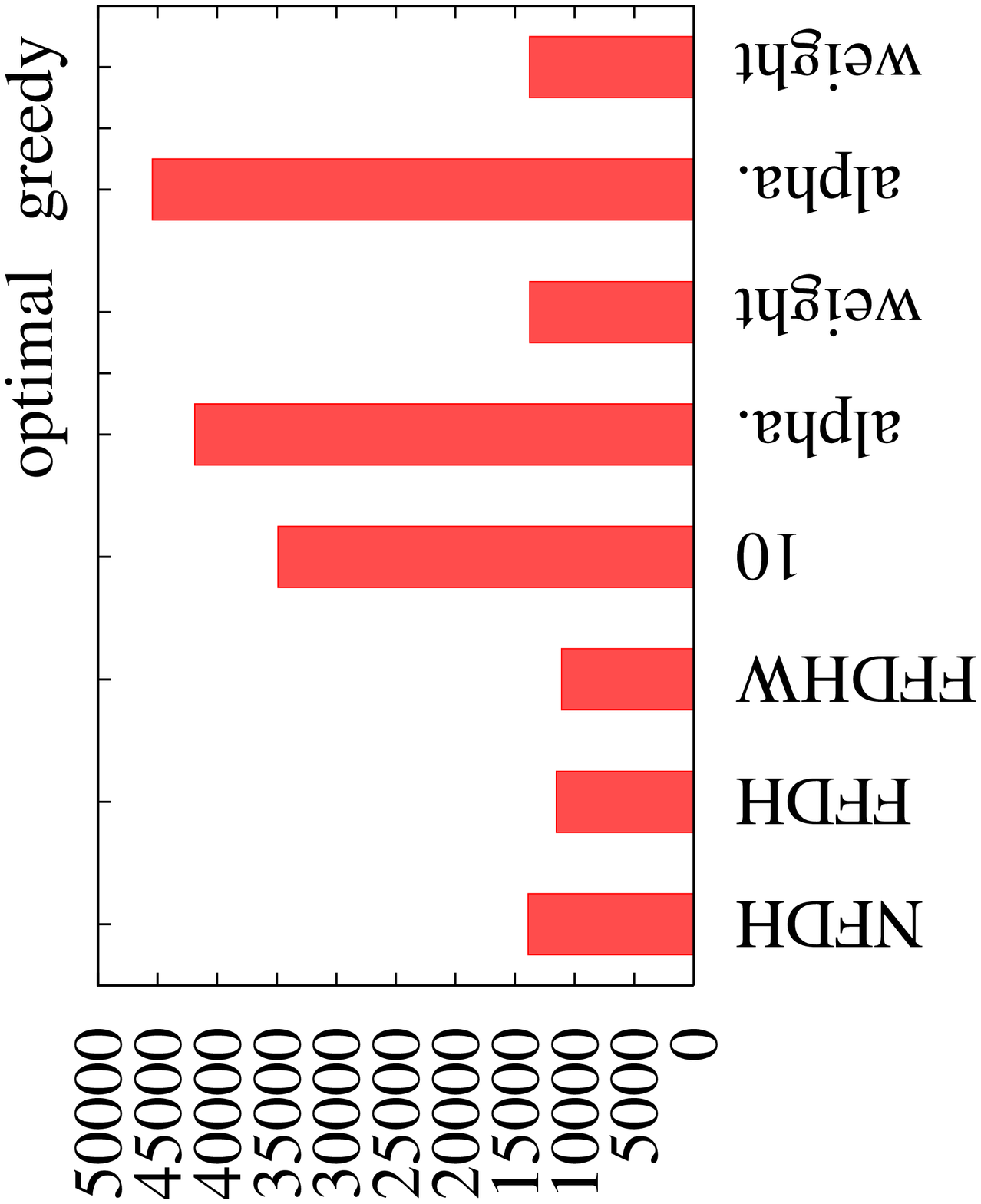}
\else
\includegraphics[height=0.47\columnwidth,angle=270]{classicresults_1_0}
\fi
}\subfigure[\label{fig:l2results}$l_2$ norm,  ZoomClouds]{
\ifpdf
\includegraphics[width=0.47\columnwidth]{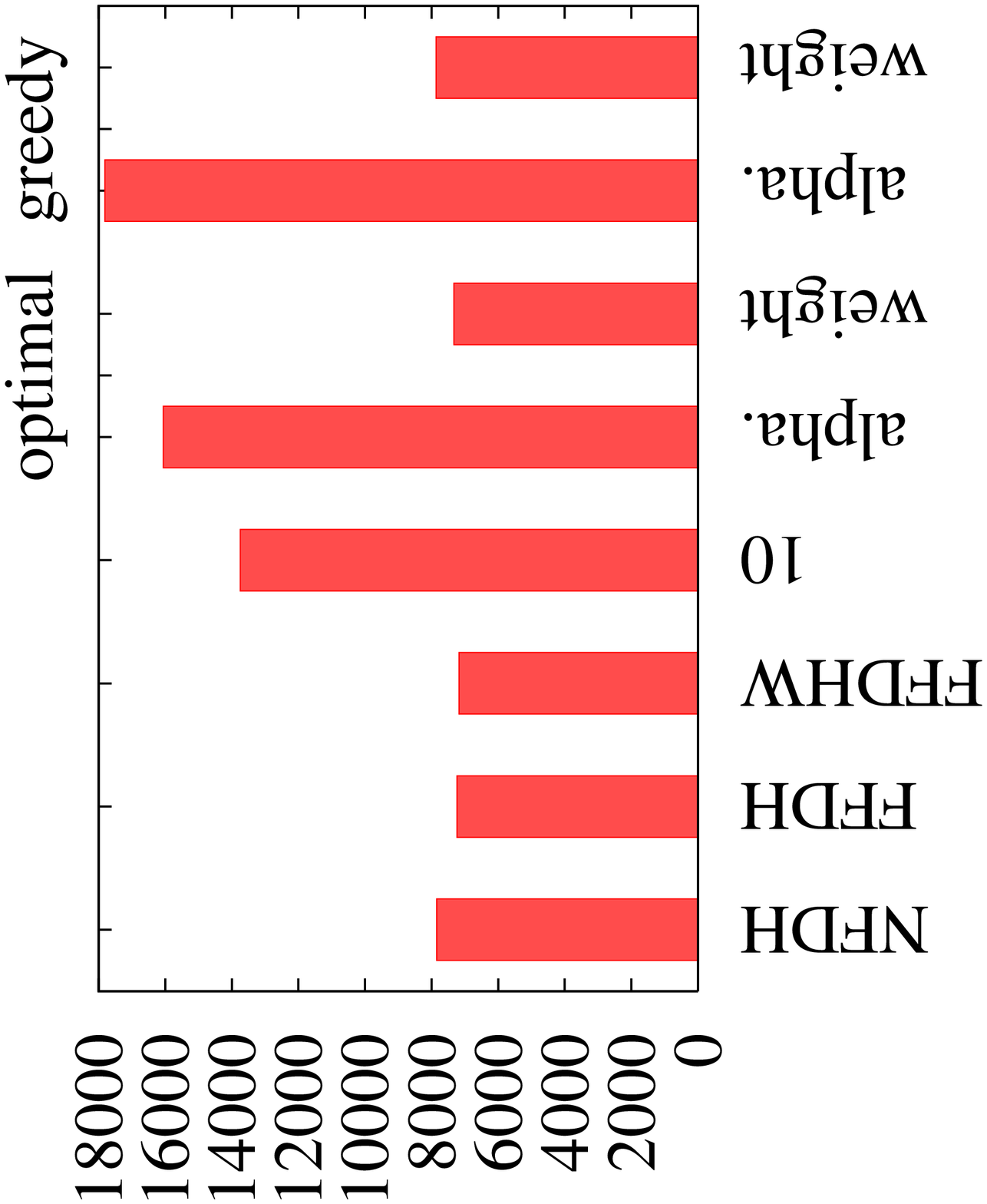}
\else
\includegraphics[height=0.47\columnwidth,angle=270]{classicresults_2_0}
\fi
}
\caption{\label{fig:algoresults}Average aggregated badnesses (in pixels), the
10-random-shuffles heuristic has the label ``10'' while the optimal
solution was computed by dynamic programming.
}
\end{figure}

The results we obtain with the badness measures are similar (see Fig.~\ref{fig:algoresults}).
The most competitive algorithms are FFDH, FFDHW and either the greedy or dynamic-programming algorithms 
applied to weight-sorted tags. The random-shuffles or
alphabetically-sorted-tags algorithms are not competitive. When considering the
$l_1$~norm of the line badnesses, the FFDH and FFDHW improve over the weight-sorted
algorithms by 24\% for the ZoomClouds data set, and by 11\% and 15\% respectively
for the Project Gutenberg data set.
When considering the $l_2$ norm, the main difference is that dynamic programming
suddenly improves over the greedy algorithm (for weight-sorted tags) by 7\%
whereas  FFDH and FFDHW only manage to improve over dynamic programming
by 1\% or 2\%. In short, if the $l_1$~norm is chosen, the FFDHW heuristic is 
the clear winner and dynamic programming is not worth the effort,
 whereas if the sum of squares is preferred, it is a close race, with
dynamic programming applied to weight-sorted tags a competitive solution.

\subsection{Tag Clouds with Arbitrary Placement}

We began with a simple EDA tool, a straightforward min-cut floorplanner
previously implemented in C by one of the authors.  The floorplanner used
techniques described by various 
authors~\cite{leng:combinatorial,leng:robust,mcfa:evaluation,zimm:estimation}.
Various modifications addressed the concerns of Sect.~\ref{placementdiffs}.

We first modified our program to perform graph
bipartitioning rather than hypergraph partitioning.
With 12 or fewer tags, bipartitioning is
done exhaustively.  The two parts must be somewhat balanced in their
total tag areas: the larger part's tag area may be at most twice the
tag area of the smaller part.  Subject to this constraint, the total
weight of edges crossing the partition is minimized.  With more than
12 tags, the system uses the Fiduccia-Mattheyses
heuristic~\cite{fidu:FM}, taking the best result of 10 runs, each
starting with a different random bipartition.  With these larger
problems, we require more balanced partitions. Again,
balance is based on total tag area, but with the constraint
that the size difference must be no larger than the area of the
largest tag in the set being bipartitioned.

The original floorplanner assumed a routing model where routing area
needs to be reserved in the areas around each
cell~\cite{zimm:estimation}.  Estimating the correct amount of
``padding'' area is not required for tag placement.
We replaced this complex code by an
estimate that a 2-pixel horizontal space
was required on the left sides of a tag, except
where the tag was on the left edge of the layout area.

The original floorplanner also preferred square layouts.
However, for tag clouds, we
have a fixed width bound that should not be exceeded.  This bias
affected the cut orientations chosen for the slicing tree, since
during placement simple heuristics monitor the estimated aspect ratios of
each floorplan area.  Originally, areas were divided by vertical cuts
when they were wider than they were high.  This is a relative decision
and does not enforce an absolute width bound. Hence, we
added an estimate of the absolute width of a floorplan area. 
Comparing this estimate against the widths of the tags for that area, we
may determine that, despite a possibly non-square floorplan area, a
vertical cut cannot be permitted.  Now, for large clouds,
near the roots of our slicing trees there continues to be
frequent switching between horizontal and vertical cuts.  However,
near the leaves, horizontal cuts predominate.

\begin{figure}
\includegraphics[width=\columnwidth]{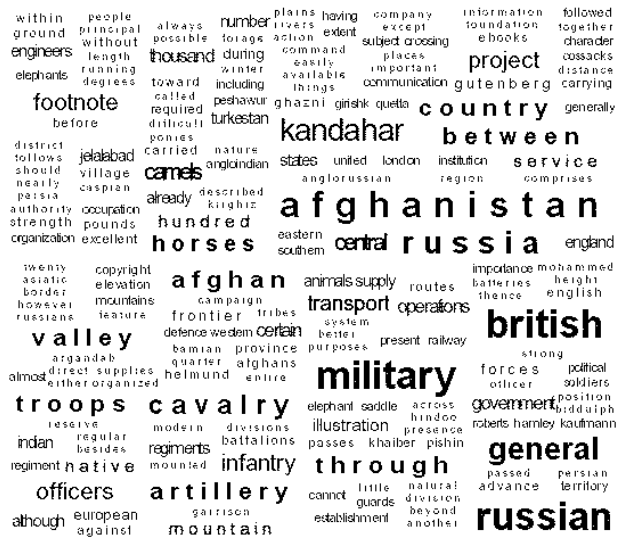}
\caption{\label{anglorussian-cloud} Large tag cloud generated from a
Project Gutenberg e-text.}
\end{figure}

Fig.~\ref{caxtons} shows an example,
a 20-tag cloud that we obtain from Bulwer-Lytton's
\textit{The Caxtons}~\cite[etext~7605]{gutenberg}.
The word `father' has been chosen
in a shorter/wider variation, whereas the word `before' was chosen to have
a taller but narrower shape than the default.  Although it may not be possible
to read the smaller tags in the  200-tag cloud
(Fig.~\ref{anglorussian-cloud})  from Rodenbough's 1885
text on the Anglo-Russian dispute~\cite[etext~7320]{gutenberg},
the cloud
shows the effect of tag resizing (`afghan\-i\-stan' being vertically compressed
and `kandahar' being horizontally compressed).  The organization of the
smaller tags into (local) columns also confirms that horizontal cut lines
predominate near the slicing-tree leaves.

\subsubsection{Results}

\begin{table}
\begin{center}
\begin{tabular}{r|rrrrr}\hline
No. Tags & Iters & Finish & Size & C-hard & C-soft\\\hline
20 &   3.0 &  38.5 &   1.0 &   24.5 &  339.0 \\
50 &   2.2 &  43.1 &   0.9 &   38.0 &  865.5 \\
100 &   1.1 &  39.2 &   1.1 &   39.0 & 1765.0 \\
200 &   1.0 &  64.6 &   1.4 &   59.0 & 6429.5 \\
\hline
\end{tabular}
\end{center}
\caption{\label{results-floorp-time}Times
(ms)  to make and size the slicing tree.
Times for block-packer { \sc compaSS} are also shown.}
\end{table}

Results are shown in Tables~\ref{results-floorp-time}--\ref{results-floorp-wirelen}.   
In the first table, the `Finish' column gives the over-all time, in
milliseconds on a 1.7\,GHz, Pentium~4-based machine.  Interestingly, the
results were obtained faster for 100-tag clouds than for 50-tag clouds:  a
tuning parameter was used in the modifications made so that the floorplanner
would not exceed a 550~pixel width.   If set incorrectly, the placement
can be much narrower, or perhaps wider, than 550~pixels.  If this is detected,
the program adjusts the parameter and tries again.  For the smallest tags,
an average of 3~attempts per cloud were made (column `Iters').
From the
`Size' column, we see that floorplan sizing was only a small part of the
over-all time.  For comparison, we ran the block-packing program
\textsc{compaSS}~\cite{chan:blobb-and-compass} against our data.  It does 
not consider tag proximity, but simply seeks a tight layout.
The `C-hard' column uses only the normal sizes of tags, whereas the
`C-soft' column shows that unacceptably long runtimes are required for
the resizing variant, where tag areas are fixed but each tag's 
aspect ratio is continuously variable
over a range. (\textsc{compaSS} was fast when 
when given a set of 3~aspect variations per tag.  
Unfortunately, 
it assumes exactly identical area for each variation.
However, with indirect control over how the browser renders
the tag variations, the three areas are not quite identical.  
Thus, we cannot compare solution qualities fairly.)

The solution area was examined in Table~\ref{results-floorp-area},
and it was compared against two row-based tag layouts: first, when
the tags were given in descending order (by height); second,
when the tags were randomly ordered.  The last column shows the
solution obtained by \textsc{compaSS}.

\begin{table}
\begin{center}
\begin{tabular}{l|rrrr}\hline
 & & \multicolumn{2}{c}{Greedy}& \\
No. Tags & Min-cut & (sorted) & (random) & \textsc{compaSS} \\\hline
20  & 31 & 37 & 46 & \textbf{29}\\
50  & 63 & 62 & 85 & \textbf{59}\\
100  & 111 & 99 & 139 & \textbf{98}\\
200  & 192 & \textbf{165} & 231 & 170\\
\hline
\end{tabular}
\end{center}
\caption{\label{results-floorp-area} Average area (kilopixels) 
for the bounding boxes of tag clouds. } 
\end{table}

The sorted greedy
heuristic used 2--19\% less area than our min-cut heuristic (although
the random greedy heuristic used 20--48\% more area than ours).
Compared with \textsc{compaSS}, min-cut used 7--13\% more area.  
These results are not surprising: \textsc{compaSS} and the
sorted greedy heuristic seek only a tight packing,
whereas min-cut also seeks to group together strongly related tags.  With 200 tags,
it is remarkable that the more sophisticated \textsc{compaSS} approach
was not as good as the greedy heuristic.  This might be attributed to
special characteristics of our data, such as the non-squareness of most
tags. For the greedy approaches and \textsc{compaSS}, only the default
shape of each tag was used. When tags were instead available with continuously
variable aspect ratios (over the range used by the 3 choices available
to our min-cut program), \textsc{compaSS} was able to reduce area by
approximately 12\%, compared with its performance when only the default
shape was allowed. (However, Table~\ref{results-floorp-time} shows that 
this required more than 6\,s on large clouds.)

The min-cut approach clearly (and unsurprisingly) outperformed 
greedy approaches and \textsc{compaSS}
when we tested proximity for semantically related tags 
(see Table~\ref{results-floorp-wirelen}).  It
considers this factor, unlike the others.

Although \textsc{compaSS} and the sorted greedy approach both 
packed tightly,  and although both are oblivious
to tag relationships, \textsc{compaSS} is apparently better at grouping
than the sorted greedy heuristic. This is counterintuitive and
reveals a weakness in using
Equation~\ref{wgtd-wirelen} to assess grouping: a tight, square packing
will score better than a loose or rectangular packing.  It appears that 
\textsc{compaSS} often uses far less than its maximum 550~pixel width.  By
nature, the greedy approach leads to widths of almost 550~pixels;
hence, its small clouds have large aspect ratios.
On small clouds, our min-cut floorplanner seeks a square
layout, assuming that each tag is itself approximately square.
The effect is that small min-cut clouds tend to have aspect ratios
similar to a typical tag: in other words, their aspect ratios often 
lie between
\textsc{compaSS}-produced clouds and clouds produced by the greedy 
heuristic.  With more tags, the width bound begins to affect all
heuristics similarly, so the effects due to aspect-ratio differences
are reduced.  

\begin{table}
\begin{center}
\begin{tabular}{l|rrrr}\hline
 & & \multicolumn{2}{c}{Greedy} &\\
No. Tags & Min-cut & (sorted) & (random) &\textsc{compaSS}\\\hline
20  & \textbf{61} & 124 & 120 & 65\\
50  & \textbf{166} & 282 & 271 & 180\\
100  & \textbf{296} & 465 & 482 & 382\\
200  & \textbf{438} & 693 & 765 & 654\\
\hline
\end{tabular}
\end{center}
\caption{\label{results-floorp-wirelen} Average total weighted distance
($\times 10^3$) using Equation~\ref{wgtd-wirelen}.
Distances were computed between the lower-left corners of tags.
}
\end{table}

\section{Conclusions}

Future work should include browser-based implementations.
For in-line text, our cloud-badness model is probably incomplete
since it ignores some basic symmetry
issues: some lines may only have a few
short tags, whereas taller lines are densely packed. It is also incomplete
because it does not take into account tag similarities, but it is not 
necessarily easy to take existing tag clouds and infer an interesting
similarity measure between tags.
Tag-cloud coloring is also open to optimization. 

Despite the differences between tag-cloud layout and
EDA placement, we plan to test an
industrial strength min-cut placement tool such as 
Capo~\cite{DBLP:journals/tcad/RoyAPM06}, to see how well it 
places tags.  However, a better metric is needed for assessing
clustering of related tags, and optimizing according to
some new metric would likely  require substantial changes to an 
existing EDA tool. 

HTML and its presentation counterpart, CSS, will probably never
directly account for representations such as tag clouds. However,
CSS3~\cite{css3-text} may introduce some new instructions which may
alleviate some problems.  For example, while it is possible to justify
an inline tag cloud with the \texttt{text-align} property, the last
line is typically not justified, a limitation addressed by the
upcoming \texttt{text-align-last} property. Also, the new
\texttt{hyphenate} property might encourage the
use of slightly more sophisticated line-breaking algorithms in browsers.

\section{Acknowledgments}  

The first author was supported in part by NSERC grant 155967, and the
second author was supported in part by NSERC grant 261437 and FQRNT grant 112381.

\bibliographystyle{abbrv}


%
\balancecolumns 
\end{document}